\documentclass[prb,twocolumn,showpacs,preprintnumbers,amsmath,amssymb,superscriptaddress,longbibliography]{revtex4-1}
\usepackage{dcolumn}
\usepackage{bm,graphicx}
\usepackage{color}
\usepackage{soul}
\usepackage{xcolor}

\begin{document}

\title{Suppressed Auger scattering and tunable light emission of Landau-quantized\\ massless Kane electrons}

\author{D.~B.~But}
\affiliation{Laboratoire Charles Coulomb, UMR CNRS 5221, University of Montpellier, Montpellier 34095, France}
\affiliation{International Research Centre CENTERA, Institute of High Pressure Physics, Polish Academy of Sciences, 01-142 Warsaw, Poland}

\author{M.~Mittendorff}
\affiliation{Helmholtz-Zentrum Dresden-Rossendorf, PO Box 510119, 01314 Dresden, Germany}
\affiliation{Universit\"{a}t Duisburg-Essen, Fakult\"{a}t f\"{u}r Physik, 47057 Duisburg, Germany}

\author{C.~Consejo}
\affiliation{Laboratoire Charles Coulomb, UMR CNRS 5221, University of Montpellier, Montpellier 34095, France}

\author{F.~Teppe}
\affiliation{Laboratoire Charles Coulomb, UMR CNRS 5221, University of Montpellier, Montpellier 34095, France}

\author{N.~N.~Mikhailov}
\affiliation{A.V. Rzhanov Institute of Semiconductor Physics,
Siberian Branch, Russian Academy of Sciences, Novosibirsk 630090,
Russia}

\author{S.~A.~Dvoretskii}
\affiliation{A.V. Rzhanov Institute of Semiconductor Physics,
Siberian Branch, Russian Academy of Sciences, Novosibirsk 630090,
Russia}

\author{C.~Faugeras}
\affiliation{Laboratoire National des Champs Magn\'etiques
Intenses, CNRS-UGA-UPS-INSA-EMFL, 25 rue des Martyrs, 38042 Grenoble, France}

\author{S.~Winnerl}
\affiliation{Helmholtz-Zentrum Dresden-Rossendorf, PO Box 510119, 01314 Dresden, Germany}

\author{M.~Helm}
\affiliation{Helmholtz-Zentrum Dresden-Rossendorf, PO Box 510119, 01314 Dresden, Germany}

\author{W.~Knap}
\affiliation{Laboratoire Charles Coulomb, UMR CNRS 5221, University of Montpellier, Montpellier 34095, France}
\affiliation{International Research Centre CENTERA, Institute of High Pressure Physics, Polish Academy of Sciences, 01-142 Warsaw, Poland}

\author{M.~Potemski}
\affiliation{Laboratoire National des Champs Magn\'etiques
Intenses, CNRS-UGA-UPS-INSA-EMFL, 25 rue des Martyrs, 38042 Grenoble, France}
\affiliation{ Faculty of Physics, Institute of Experimental Physics, University of Warsaw, ul. Pasteura 5, 02-093 Warszawa, Poland}

\author{M.~Orlita}\email{milan.orlita@lncmi.cnrs.fr}
\affiliation{Laboratoire National des Champs Magn\'etiques
Intenses, CNRS-UGA-UPS-INSA-EMFL, 25 rue des Martyrs, 38042 Grenoble, France}
\affiliation{Charles University, Faculty of Mathematics and Physics, Ke Karlovu 5, 121 16 Prague 2, Czech Republic}


\maketitle

{\bf The Landau level laser has been proposed a long time ago as a unique source of monochromatic radiation, widely tunable
in the THz and infrared spectral ranges using a magnetic field. Despite many efforts,
this appealing concept never resulted in the design of a reliable device. This is due to efficient Auger scattering of
Landau-quantized electrons, which is an intrinsic non-radiative recombination channel that eventually gains over cyclotron
emission in all materials studied so far: in conventional semiconductors with parabolic bands, but also in graphene with massless electrons.
Auger processes are favored in these systems by Landau levels (or their subsets) equally spaced in energy.
Here we show that this scheme does not apply to massless Kane electrons in gapless HgCdTe, where undesirable Auger scattering
is strongly suppressed and sizeable cyclotron emission observed, for the first time from massless particles.
The gapless HgCdTe thus appears as a material of choice for future Landau level lasers.}
\vspace{0.0mm}

When a magnetic field is applied to a solid, the continuous density of electronic states transforms into a set of
discrete energy levels, known as Landau levels (LLs). Electrons excited in such a ladder may recombine
with emission of photons. This process can be viewed as an inverse of cyclotron resonance~\cite{CohenAIP05} and it is
referred to as cyclotron emission\cite{Lax60,Gornik80,GornikPBC84,KnapRSI92}. The idea to construct the LL laser by achieving stimulated cyclotron emission\cite{Lax60}
is as old as the experimental realization of the very first laser itself\cite{MaimanNature60}.
Wavelength tunability represents a great advantage of this concept. The strength of the magnetic field $B$ defines the spacing between LLs, and therefore,
also the emission frequency of the laser. This frequency typically corresponds
to the far infrared (terahertz) spectral range. The successful realization of the LL laser\cite{Lax60,WolffPPF64,WolffUS66,AokiAPL86,MorimotoPRB08} would thus bridge
the so-called terahertz gap\cite{SirtoriNature02,TonouchiNaturePhoton07}, which still exists despite a considerable effort of several generations of physicists.

There is, however,  a fundamental difficulty that has been recognized during the very first attempts to test the concept of a LL laser.
In conventional materials, with quadratically dispersing electronic bands, and therefore, with a well-defined mass $m$ of charge carriers,
the LL spectrum is equidistant: $E^S_N=\hbar \omega_c(N+1/2)$ for $N=0,1,2$, with the spacing defined by the classical cyclotron
frequency $\omega_c=eB/m$. In such materials, cyclotron emission is an intrinsically weak effect~\cite{Lax60,GornikPRL78,Gornik80}. This is due to two independent processes
which compete with cyclotron emission: reabsorption of cyclotron radiation\cite{Lax60} and Auger inter-LL recombination\cite{PotemskiPRL91,TsitsishviliPRB97} (Fig.~1a).
At higher pumping rates, the latter becomes extremely efficient for electrons within the equidistant LL ladder, and in consequence,
the excited LLs become depleted with a characteristic rate proportional to
the number of electrons in this level ($\tau^{-1}\propto n$)\cite{GornikPRL78}. Stronger pumping, either optical or electrical,
in systems with parabolic bands does not enhance cyclotron emission, but instead, increases the probability of Auger scattering.
The inversion of population thus cannot be achieved. Electrons promoted via Auger processes to higher LLs then relax
via non-radiative channels, most often by emission of optical phonons\cite{Gornik80}.

\begin{figure*}[t]
      \includegraphics[width=1\textwidth]{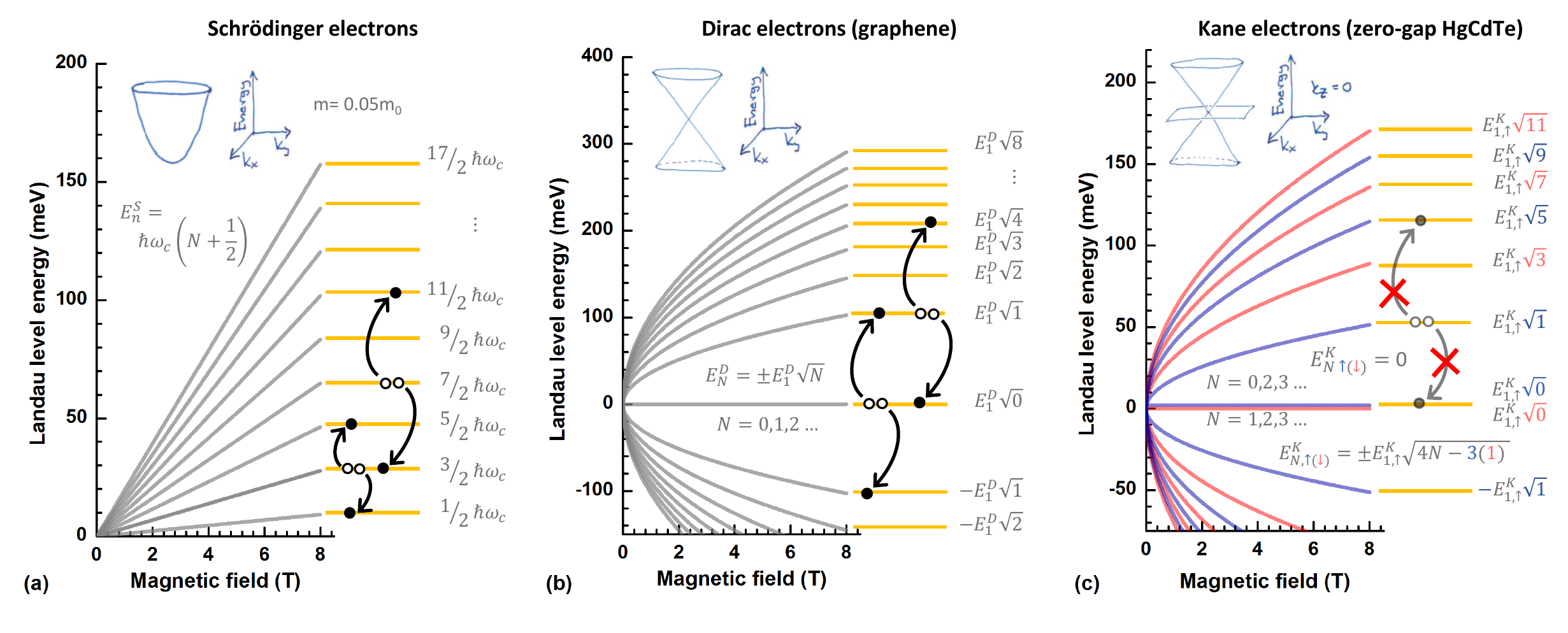}
      \caption{\label{LLs} \textbf{Landau level ladders and possible paths for Auger scattering of massive and massless electrons.} The $B$-dependence and spacing of Landau levels for conventional massive (Schr\"{o}dinger) electrons, Dirac electrons in graphene and massless Kane electrons (at $k=0$), in part (a), (b) and (c), respectively. The arrows in LL ladders correspond to the Auger scattering processes discussed in the main text.}
\end{figure*}

To overcome this obstacle, various systems with non-equidistant LLs have been proposed
as appropriate materials for a LL laser\cite{SchneiderPRl59,Lax60}. These may be found, for instance, in different narrow-gap materials or in the valence band of zinc-blende semiconductors.
So far, stimulated cyclotron emission was achieved only in bulk germanium in which a non-equidistant spacing of valence-band LLs
was created by crossed electric and magnetic fields\cite{UnterrainerPRL90},
alas in the regime close to the electrical breakdown of this material. In other attempts, quantum cascade structures have also been tested
to achieve efficient cyclotron emission\cite{BlaserAPL02,JasnotAPL12}.

Renewed impetus to investigate cyclotron emission came along with the fabrication of graphene, which was considered as an ideal system with non-equidistant LLs
for the implementation of the LL laser\cite{MorimotoPRB08,MorimotoJoPCF09}. Its conical band may be viewed as an extreme case of non-parabolicity
and implies, for a given velocity parameter $v$, a specific sequence of LLs, $E^D_N=\pm v \sqrt{2e\hbar B N}$ for $N=0,1,2\ldots$, which
displays the $\sqrt{N}$ spacing, typical of massless Dirac electrons (Fig.~1b). However, to the best of our knowledge, no cyclotron emission, let alone the Landau level laser based on graphene,
has been reported so far, despite pertinent theoretical predictions and numerous
experimental attempts\cite{MorimotoPRB08,MorimotoJoPCF09,WendlerSR15,WangPRA15,BremPRB17,ColeIEEE17,BremPRM18,PlochockaPRB09}.

This surprising lack of optical emission seems to be still a consequence of Auger scattering. Initially, such processes were expected to vanish in
Landau-quantized graphene, but finally they were experimentally proven to be present and very efficient\cite{MittendorffNaturePhys15,Konig-OttoNL17}.
The reason behind is simple. The LL spectrum in graphene: $E^D_N\propto \sqrt{N}$ for $N=0,1,2 \ldots$, includes subsets of
equidistant levels. For instance, LLs with indices $N=0, \pm1, \pm4, \pm9\ldots$
or $0, \pm2, \pm8, \pm18\ldots$, are equally spaced and may thus support, similar to conventional materials,
efficient Auger recombination (Fig.~1b). Notably, such a conclusion is valid not only for graphene, but for any other system with a conical band
described by the Dirac Hamiltonian for particles with a zero rest mass.

\begin{figure*}[t]
      \includegraphics[width=.88\textwidth]{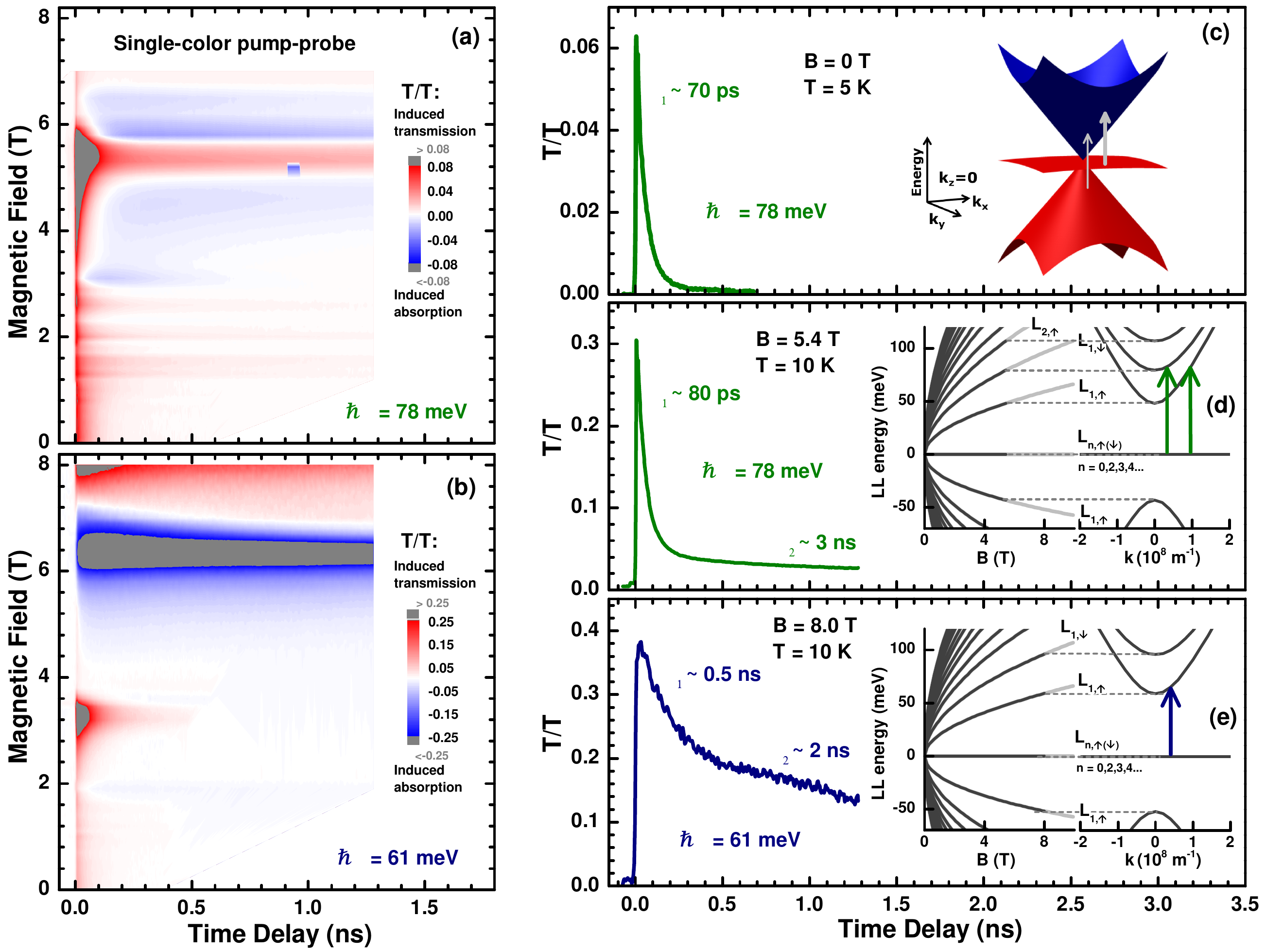}
      \caption{\label{Pump-probe} \textbf{Pump-probe experiments and Auger scattering of Landau-quantized massless Kane electrons.} (a) and (b) False-color maps of induced transmission $\Delta T/T$ measured
      at $T=10$~K and for laser photon energies $\hbar\omega=78$~meV (upper panel) and 61~meV (lower panel) as a function of
      the time delay and magnetic field, showing both, pronounced induced transmission ($\Delta T/T>0$) and absorption ($\Delta T/T<0$).
      (c-e) Selected pump-probe transients. The inter-LL excitations in (d) and (e) follow the standard selection rules for electric-dipole transitions which are active in linearly polarized light, $N\rightarrow N\pm1$, and schematically shown in the corresponding insets.}
\end{figure*}

Fortunately, one may find other systems with massless electrons, described by a Hamiltonian that differs from the Dirac one.
In such systems, equidistantly spaced LLs may be avoided and undesirable Auger scattering possibly suppressed.
In this paper,
we test Landau-quantized 3D massless Kane electrons\cite{KaneJPCS57,Kacmanppsb71,OrlitaNaturePhys14,TeppeNatureComm16} against the efficiency
of Auger scattering and against the possibility of efficient LL emission. This type of massless electrons appears in conventional bulk zinc-blende
semiconductors when their energy band gap -- the parameter responsible for the generation of the non-zero band mass of electrons and holes, and therefore, also
Schr\"{o}dinger-like behavior of these carriers -- vanishes, for instance, in the ternary compound HgCdTe explored in this paper.

In magnetic fields, the conical band of 3D massless Kane electrons splits into LLs, or more precisely, Landau bands,
which resemble those of massless Dirac electrons\cite{Berestetskii}:
\begin{equation}\label{Kane_LLs}
E^K_{N,\sigma}(k)= \pm v \sqrt{2e\hbar B(N-1/2+\sigma/2)+\hbar^2 k^2},
\end{equation}
where $N=1,2,3\ldots$ is the LL index and $k$ the momentum along the applied magnetic field. However, the splitting due to spin
is given by $\sigma=\pm 1/2$ and differs from the one known for genuine Dirac-type electrons ($\sigma=\pm 1$).
As a result, the LL spectrum of massless Kane electrons (Fig.~1c): $E^K_N\propto \sqrt{N}$ for $N=0,1,3,5\ldots$, does not include
any subsets of equally spaced LLs at $k=0$, where the maxima in the density of states are. In contrast, 3D Dirac electrons with $\sigma=\pm 1$
display at $k=0$ the spacing $E^D_N\propto \sqrt{N}$ where $N=0,1,2,3\ldots$, which is identical to that of 2D graphene.

Moreover, there exists a set of narrowly spaced LLs, which originates from the weakly dispersing (nearly flat) band.
In the simplest approach, these LLs do not disperse with $B$: $E^K_{n,\sigma}=0$ for $n=0,2,3,4\ldots$.
Excitations from this flat band to conduction-band LLs constitute a dominant contribution
to the optical response at low photon energies and follow the standard selection rules, $N\rightarrow N\pm1$, for electric-dipole transitions\cite{WeilerSS81,OrlitaNaturePhys14,TeppeNatureComm16}.

To test Landau-quantized electrons in gapless HgCdTe for the presence of Auger scattering, we have employed a degenerate pump-probe technique, using the
free-electron laser emitting $\sim$3-ps-long pulses in the mid-infrared range. The results are plotted in Figs.~2a and b,
in the form of false color-maps, for two different photon energies of the laser. These maps show pump-probe transients, \emph{i.e.}, the pump-induced relative
change of transmission, $\Delta T/T$, as a function of the applied magnetic field and time delay between pump and probe pulses. One immediately
concludes that $\Delta T/T$ undergoes a fairly complex evolution in both, its amplitude and sign, on a characteristic time scale extending up to nanoseconds.

To illustrate the observed behavior in greater detail, three selected pump-probe traces are plotted in Figs.~2c-e. At $B=0$, the observed induced transmission (bleaching) is primarily due to electrons excited from the fully occupied heavy-hole-like flat band to the upper conical band (inset of Fig.~2c). The contribution of excitations from the lower cone is considerably weaker. The induced transmission exhibits a nearly mono-exponential decay, with a characteristic time, $\tau=70$~ps, which is comparable to that in other gapless systems with conical bands, such as graphene\cite{WinnerlPRL11}. This relatively fast relaxation may be associated with the emission of phonons.

When the magnetic field is applied, one observes a pronounced change in $\Delta T/T$ transients. For particular values of $B$, the second component develops, with a characteristic decay time in the nanosecond range. The appearance of this slow component clearly coincides with the resonant pumping into the bottom of Landau bands. This is illustrated in Fig.~2d,
where the $\Delta T/T$ trace collected at $B=5.4$~T and $\hbar\omega=78$~meV is plotted. In this particular case, electrons are pumped from the flat band to two lowest lying
conduction-band LLs, see the inset of Fig.~2d. The faster component, with a relaxation rate not much different from the zero-field one ($\tau\approx 80$~ps), can be associated with the relaxation of electrons excited into the lowest Landau band ($E^K_{1,\uparrow}$).
This relaxation of electrons from states with finite-momentum towards $k\approx0$ is probably due to emission of phonons.
Most likely, electrons relax via emission of acoustic phonons. The slow component, appearing only when  electrons are pumped resonantly from/to states around $k\approx 0$, reflects
the existence of long-living electrons at the bottom of the $E^K_{1,\downarrow}$ Landau band. Another example of the slowly decaying component in the pump-probe transient is shown in Fig.~2e. This $\Delta T/T$ trace was collected at $B=8$~T and $\hbar\omega=61$~meV, when only the lowest LL in the conduction band ($E^K_{1,\uparrow}$) is pumped close to zero-momentum states.

\begin{figure}
      \includegraphics[width=0.47\textwidth]{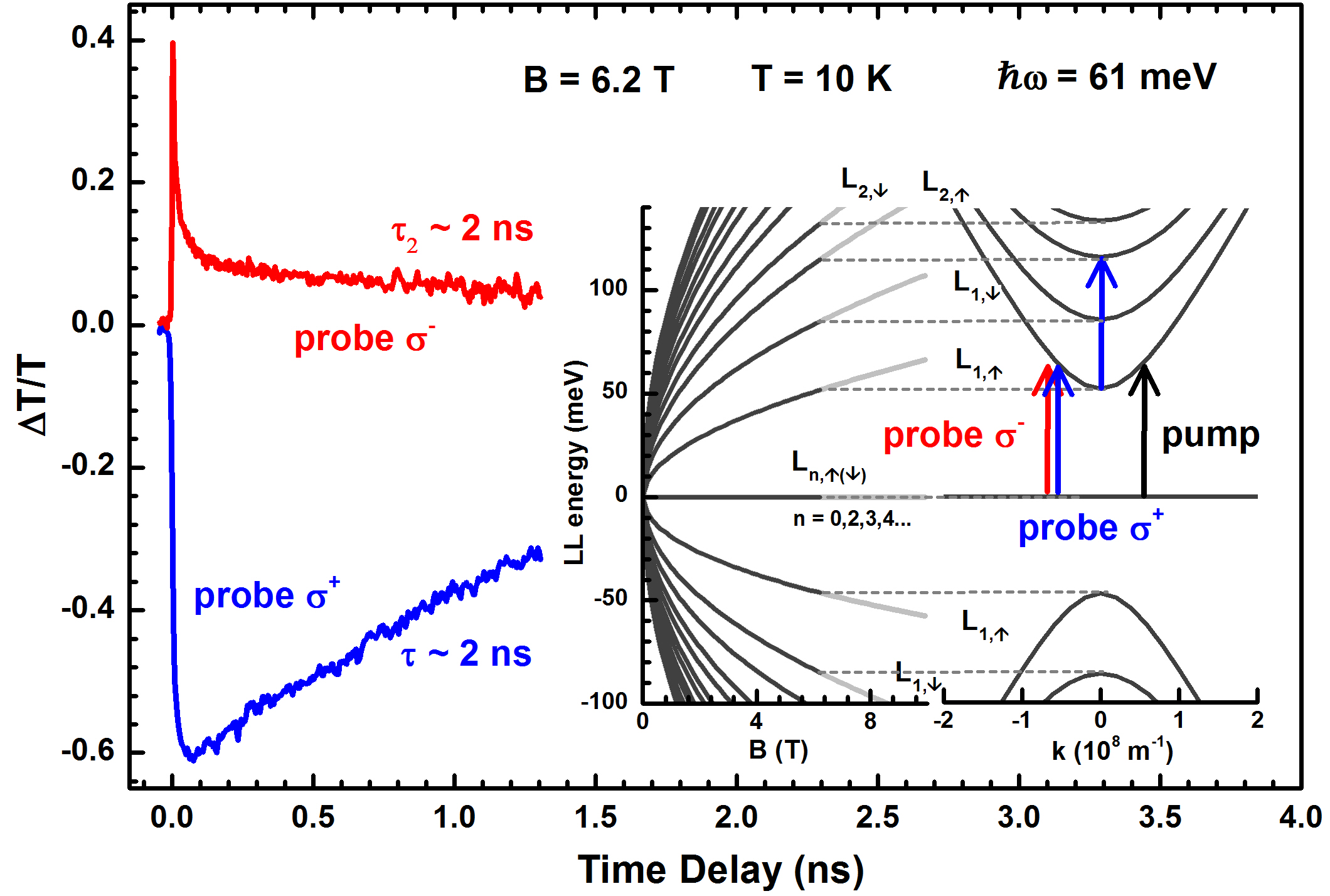}
      \caption{\label{Pump-probe-polarization-resolved} \textbf{Pump-probe experiments using circularly polarized probe beam.} Pump-probe transients collected at $B=6.2$~T using the laser photon energy of $\hbar\omega=61$~meV and circularly polarized probe beam. The involved excitations, which follow the standard electric-dipole selection rules for right and left circularly polarized radiation: $N\rightarrow N+1$ and $N\rightarrow N-1$, respectively, are schematically shown in the inset.}
\end{figure}

Experiments with circularly polarized radiation represent another way to visualize how the relaxation of photo-excited electrons in gapless HgCdTe is slowed-down when
the magnetic field is applied.
Pump-probe transients collected using a linearly polarized pump pulse, but with a circularly polarized probe pulse are depicted in Fig.~3. In this particular case, electrons are excited by the pump pulse from the flat band to the lowest lying LL only, as schematically shown by the vertical arrow in the inset of Fig.~3. The transient recorded using $\sigma^{-}$-polarized probe shows positive $\Delta T/T$, with an initial fast decay due to the relaxation of electrons to the bottom of $E^K_{1,\uparrow}$ level and slower component
indicating a nanosecond-long lifetime of excited electrons around $k\approx 0$. Different behavior, \emph{i.e.}, strong induced absorption $\Delta T/T<0$ ,  is observed when the probing beam is $\sigma^{+}$ polarized. In this latter case, the induced absorption is primarily due to the $E^K_{1,\uparrow} \rightarrow E^K_{2,\uparrow}$ transition which becomes activated due to electrons promoted to the $E^K_{1,\uparrow}$ level by the pump pulse. This excitation corresponds to the fundamental cyclotron mode, which becomes resonant with the laser photon energy ($\hbar\omega=61$~meV) just at the selected magnetic field of $B=6.2$~T.

Let us now discuss the main findings of our pump-probe experiments. These show that electrons in Landau-quantized gapless HgCdTe relax with relatively long decay times, at the scale of nanoseconds. In addition, the presented pump-probe experiments were performed at relatively high photon fluences, $\sim$0.5~$\mu$J.cm$^{-2}$. These translate, for the chosen photon energy, into the photon flux of $10^{14}$~cm$^{-2}$ per pulse. Due to the relatively large absorption coefficient\cite{OrlitaNaturePhys14}, which in the linear regime, at $\hbar\omega\approx 70$~meV and $B=0$, reaches $\lambda \approx 3\times 10^3$~cm$^{-1}$, a significant part of the flux is absorbed in the explored 3-$\mu$m-thick layer of gapless HgCdTe.
This corresponds to more than $10^{16}$~cm$^{-3}$ electrons promoted from the flat band to the upper conical band by a single pump pulse. Since there is no decay observed at the scale of the pulse duration ($\sim$3~ps), this number serves as a rough estimate of the electron density in the sample just after the pump pulse.

\begin{figure*}[t]
      \includegraphics[width=.96\textwidth]{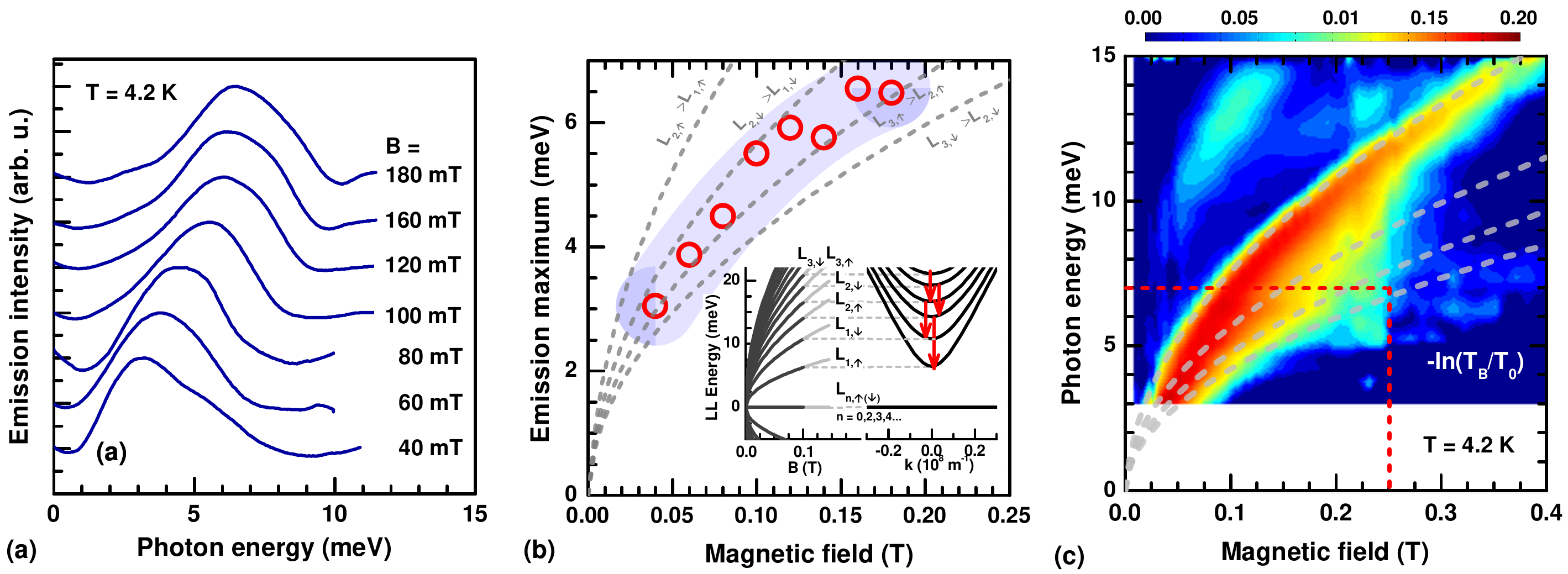}
      \caption{\label{Emission} \textbf{Cyclotron-emission and absorption of 3D massless Kane electrons.} (a) Cyclotron emission spectra measured on gapless HgCdTe sample kept in the liquid helium at selected values of the magnetic field. (b) Experimentally deduced maxima of emission compared with theoretically expected cyclotron modes (dashed lines), see the text. The grey area correspond to the
      error bar of cyclotron emission maxima. The LL spectrum with schematically depicted emission lines (by vertical arrows) is shown in the inset. (c) The false-color map of magneto-absorbance, primarily due to cyclotron resonance absorption, reprinted from Ref.~\onlinecite{OrlitaNaturePhys14}. The dashed lines show the theoretically expected
      positions of cyclotron modes, the same ones as in (b). The red dashed lines shows the range of magnetic fields and photon energies in which cyclotron emission was explored.}
\end{figure*}

One may compare this carrier density and the deduced relaxation time, with the results obtained in conventional semiconductors, where a strong decrease of relaxation
time with the electron density has been reported: $\tau \propto n^{-1}$, see Refs.~\onlinecite{GornikPRL78,Gornik80,GornikPBC84}. In conventional semiconductors, the nanosecond
decay times are only found at very low carrier densities in the excited LL, in the range of 10$^{12}$-10$^{13}$~cm$^{-3}$. For the carrier density close to $10^{16}$~cm$^{-3}$ in a parabolic-band semiconductor, one expects the electron relaxation time to drop down to the sub-picosecond range\cite{Gornik80} due to efficient Auger processes. This is more than three orders of magnitude less than relaxation times observed in gapless HgCdTe in a magnetic field. This indicates that Auger scattering is indeed strongly suppressed for Landau-quantized massless Kane electrons as compared to other materials explored so far. We interpret this suppression as a direct consequence of the specific LL spectrum, which does
not include any subset of equidistant levels (around $k\approx 0$).

Since the rate of inter-LL Auger scattering follows the degeneracy of Landau levels ($\propto B$) and the strength of interaction between Landau-quantized electrons
($\propto \sqrt{B}$)\cite{GoerbigRMP11}, we expect similarly long lifetimes also at lower magnetic fields. Let us also note that the observed lifetime of electrons, even though unusually long as compared to any so-far explored Landau-quantized semiconductor or semimetal\cite{Gornik80}, remains still significantly shorter than spontaneous cyclotron radiative lifetime. This latter time can be for massless electrons estimated using the simple formula\cite{MorimotoPRB08}, $\tau_{\mathrm{sp}}^{-1} \sim \alpha \omega_c (v/c)^2$, where $\alpha$ is the fine structure constant, $c$ speed of light in vacuum and the characteristic $\omega_c$ cyclotron frequency ($\tau_{\mathrm{sp}}\approx 10^{-6}$~s in the THz range for $v=10^6$~m/s). Hence, there still exist other (non-radiative) channels which dominate the recombination, such as electron-phonon interaction giving rise to emission of phonons.

The observed slowing-down in the relaxation dynamics of electrons induced by the magnetic field, with the overall lifetime of photo-excited electrons in the nanosecond
range, calls for cyclotron emission experiments. We have performed such experiments in the THz spectral range, which is the most relevant one for applications of the (future) Landau level laser technology. To generate cyclotron emission, the sample was placed in a superconducting coil at liquid helium temperature and electrically pumped, using ms-long current pulses, see Methods section. The emitted radiation was analyzed using a photoconductive InSb detector, with a spectrally narrow cyclotron-resonance response (with the linewidth of $\sim$1~meV)\cite{YavorskiyJAP18}, tunable by a specially dedicated coil. The collected cyclotron emission spectra are plotted in Fig.~4a for selected values of the
magnetic field applied to the sample.

A brief inspection of the emission spectra suggests that they feature a single emission band. Its position in the spectrum is tunable in the THz range
by a relatively low magnetic field (tens of millitesla) and it roughly follows a $\sqrt{B}$ dependence, which is typical of massless electrons (Fig.~4b).
This also agrees well with results of cyclotron absorption experiments performed on the same sample, see Fig.~4c and Ref.~\onlinecite{OrlitaNaturePhys14}.
However, a closer look at the lineshape implies that several emission modes actually contribute. Theoretically, one indeed expects
several cyclotron emission modes in the interval given by the linewidth (FWHM$\sim$~4~meV). In Fig.~4b, the dashed lines
show the positions of cyclotron emission modes with the final states of electrons in the four lowest lying conduction-band LLs: $E^K_{N+1,\uparrow(\downarrow)}\rightarrow E^K_{N,\uparrow(\downarrow)}$ for $N=1$ and 2. It is thus the spacing of these cyclotron modes, together with primarily elastic scattering
processes, which is responsible for the observed width of the emission band. The mutual intensities of contributing modes then determine the position of the
emission band maximum. At higher $B$, the relative weight of modes from higher LLs increases, which reflects the distribution of electrons
among LLs established by electrical pumping and which leads to a slowing-down of the $\sqrt{B}$ dependence.

Even though the cyclotron emission observed in the used configuration is primarily due to spontaneous recombination, it is worth to discuss conditions
required to obtain stimulated emission and gain. To achieve light amplification comparable to, \emph{e.g.}, quantum cascade lasers~\cite{FaistScience94,Sirtori,WilliamsNaturePhoton07}, the gain coefficient
has to approach $g=\Delta n\cdot\sigma\sim 100$~cm$^{-1}$, where $\Delta n$ stands for the population inversion and $\sigma = \lambda^2/(2\pi)\cdot(\tau_{\mathrm{tot}}/\tau_{\mathrm{sp}})$ is the stimulated emission cross-section\cite{Saleh}. The typical spontaneous cyclotron emission lifetime in the THz range
for massless electrons reaches $\tau_{\mathrm{sp}}\sim 1$~$\mu$s. The total lifetime
is defined by the width of the emission line, $\tau_{\mathrm{tot}} \sim 1$~ps, and in our case, it dominantly reflects
elastic scattering processes. Taking the characteristic wavelength of $\lambda = 300$~$\mu$m, we obtain the cross-section $\sigma \sim 10^{-10}$~cm$^{2}$. This implies the necessity to achieve the population inversion $\Delta n \gtrsim 10^{12}$~cm$^{-3}$. Our simple estimate, see Supplementary
materials\cite{SM}, indicates that such a population inversion should be achievable at least in the pulsed regime.

To conclude, we have demonstrated, for the first time, cyclotron emission of massless electrons. This emission was observed in gapless HgCdTe -- a system hosting 3D massless Kane electrons.
The existence of sizeable cyclotron emission is directly related to their particular Landau level spectrum, which comprises only non-equidistantly spaced levels. The systems hosting massless Kane electrons are thus promising candidates for an active medium of a Landau level laser, which would, in this particular case, operate in the THz and infrared spectral
ranges and would be widely tunable by very low magnetic fields.

\vspace{4mm}
{\small
\noindent\textbf{Methods}
\vspace{1mm}

\emph{Sample growth.} The sample was grown using standard molecular-beam epitaxy on a
(013)-oriented semi-insulating GaAs substrate. The growth sequence
started with ZnTe and CdTe transition regions, followed by the MCT
epilayer with gradually changing cadmium content~$x$.
The prepared MCT layer contains a region with $x\approx{0.17}$ of thickness
$d\approx{3}.2\:\mu\mbox{m}$. For more details about the explored sample see Ref.~\onlinecite{OrlitaNaturePhys14}.
\vspace{1mm}

\emph{Pump-probe spectroscopy.} The free-electron laser FELBE provided
frequency-tunable Fourier-limited radiation pulses. In the experiments
described in this paper, photon energies of $\hbar\omega=61$ and 78~meV were chosen.
The pulse duration was about 3~ps, the repetition rate was 13~MHz.
The pulses were split into pump and probe pulses by a pellicle beam splitter. The
polarizations of pump and probe beams were controlled independently.
Frequency-tunable quarter-wave plates (from Alphalas GmbH) were used for the
generation of circularly polarized radiation. Both the pump and probe beam were
focused on the sample in the magnet cryostat by an off-axis parabolic mirror
(effective focal length: 178~mm). The spot size on the sample was $\sim$~0.5~mm
(FWHM). The pump fluence was $\sim$~0.5~$\mu$J.cm$^{-2}$,
the fluence of the probe beam was about 10\% of the
pump fluence. The time delay between pump and probe pulses was varied using a
mechanical delay stage.
\vspace{1mm}

\emph{Emission measurements.} To measure radiation emitted due to inter-LL recombination
of electrons, the radiation was guided, using a copper light pipe, to a InSb photoconductive
spatially separated from the coil inside which the sample was placed. The used detector
has a narrow linewidth ($\sim$1~meV) and a field-tunable response due to cyclotron resonance absorption.\cite{YavorskiyJAP18}
This tunability is ensured by a specially dedicated superconducting coil and allows us to
analyze the radiation in the spectral range of 0.4-2.5~THz. All emission experiments
operate in a pulsed mode (using current pulses for pumping), with the pulse duration
of 7~ms and with the peak-to-peak value of electric field up to 12~V/cm. The duty circle was tuned
in such a way that the average power consumption of emitter did not exceed
13 mW (to avoid sample heating). The signal on the InSb detector was collected using a conventional lock-in technique.


\vspace{2mm}
\noindent\textbf{Acknowledgements}
\vspace{1mm}

The authors acknowledge helpful discussions with D. M. Basko and R. J. Nicholas. This work was supported by ANR DIRAC3D project.
The research  was also partially supported by the CNRS through the LIA TeraMIR project, by the Occitanie region and MIPS department of Montpellier University via the Terahertz Occitanie platform, and by the Foundation for Polish Science through the TEAM and IRA Programs financed by EU within SG OP Program.
Part of this work has been supported by the project CALIPSO under the EC Contract No. 312284.

\vspace{2mm}
\noindent\textbf{Author contributions}
\vspace{1mm}

The experiment was proposed by M.O. and M.P. The sample was grown by N.N.M. and S.A.D. Time-resolved and cw
magneto-optical experiments were carried out by M.O., M.M., S.W., C.F. and M.H. The cyclotron emission experiments were performed by
D.B., C.C., F.T. and W.K. All coauthors discussed the experimental data and interpretation of results.
M.O. and M.P. wrote the manuscript, all coauthors commented on it.

\vspace{3mm}
\noindent\textbf{Additional information}
\vspace{1mm}

The authors declare no competing financial interests. Correspondence and requests for materials should be addressed to M.O.

\vspace{2mm}
\noindent\textbf{Data availability}
\vspace{1mm}

The data that support the plots within this paper and other findings of this study are available from the corresponding author upon reasonable request.

\vspace{2mm}
\noindent\textbf{Code availability}
\vspace{1mm}

The code for modelling of cyclotron mode energies are available from the corresponding author on reasonable request.

\newpage
\pagenumbering{gobble}

\begin{figure}[htp]
\includegraphics[page=1,trim = 17mm 17mm 17mm 17mm, width=1.0\textwidth,height=1.0\textheight]{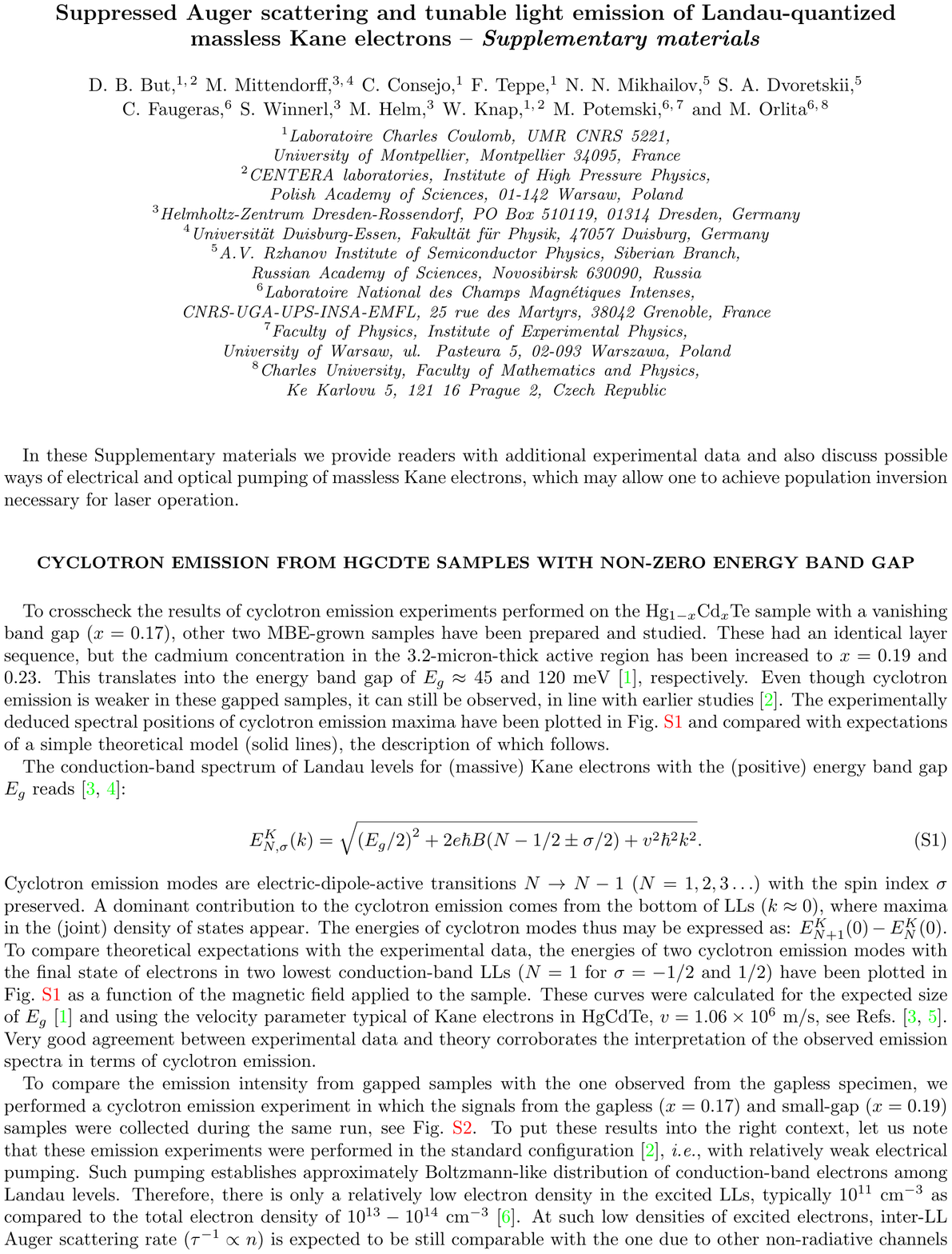}
\end{figure}

\newpage

\begin{figure}[htp]
  \includegraphics[page=2,trim = 17mm 17mm 17mm 17mm, width=1.0\textwidth,height=1.0\textheight]{SI.pdf}
\end{figure}

\newpage

\begin{figure}[htp]
  \includegraphics[page=3,trim = 17mm 17mm 17mm 17mm, width=1.0\textwidth,height=1.0\textheight]{SI.pdf}
\end{figure}

\newpage

\begin{figure}[htp]
  \includegraphics[page=4,trim = 17mm 17mm 17mm 17mm, width=1.0\textwidth,height=1.0\textheight]{SI.pdf}
\end{figure}

\end{document}